\documentclass{aa}

\usepackage{graphicx}
\usepackage{natbib}
\usepackage{txfonts}

\begin{document}

\title {Abundance gradients in the Milky Way for $\alpha$ elements, Iron peak elements,
 Barium, Lanthanum and Europium}

\author {G. Cescutti\inst{1}
\thanks {email to: cescutti@oats.inaf.it}
\and  F. Matteucci\inst{1, 2}
\and  P. Fran\c cois\inst{3,4}
\and  C. Chiappini \inst{2}}

\institute{ Dipartimento di Astronomia, Universit\'a di Trieste, via G.B. Tiepolo 11, I-34131  
\and  I.N.A.F. Osservatorio Astronomico di Trieste, via G.B. Tiepolo 11, I-34131
\and  Observatoire de Paris/Meudon, GEPI, 61 Avenue de l'Observatoire, 75014 Paris, France
\and  European Southern Observatory, Casilla 19001, Santiago 19, Chile}
\date{Received xxxx / Accepted xxxx}

\abstract{}{To model the abundance gradients in the disk  of the Milky Way for 
several chemical elements (O, Mg, Si, S, Ca, Sc, Ti, Co, V, Fe, Ni, Zn, Cu, Mn, Cr, Ba, La and Eu), and
compare our results with the most recent and  homogeneous observational data.}
{We adopt a chemical evolution model able to well reproduce the main properties
of the solar vicinity. The model assumes that the disk formed inside-out
with a timescale for the formation of the thin disk of 7 Gyr in the solar vicinity, 
whereas the halo formed on a timescale of 0.8 Gyr. We also adopt new empirical stellar yields 
derived to best fit the abundances and the abundance ratios of the solar vicinity.}
{We compute, 
for the first time, the abundance gradients for all the above mentioned 
elements in the galactocentric distance
range 4 - 22 kpc. The comparison with the observed data on Cepheids in the galactocentric distance range 
5-17 kpc gives 
a very good agreement for many of the studied elements.
In addition, we fit very well the data for the evolution
of Lanthanum in the solar vicinity for which we present results here for the first time. 
We explore, also for the first time, the behaviour of 
the abundance gradients at large galactocentric distances by comparing our results with data 
relative to distant
 open clusters and red giants and select the best chemical evolution model model on the basis of that.}
{We find a very good fit to the observed abundance gradients, as traced by Cepheids, for most of
the elements, thus confirming the validity of the inside-out scenario for the formation 
of the Milky Way disk as well as the adopted nucleosynthesis prescriptions. }


\titlerunning{Abundance gradients in the MW}

\maketitle

\authorrunning{Cescutti et al.}

\section{Introduction}

Understanding the formation and the evolution of the Milky Way is 
fundamental to improve the knowledge of the formation of spiral galaxies in general.
Many are the observational constraints in the Milky Way.
 The most important ones are represented by the evolution of the abundances of the
chemical  elements.
Recently, many chemical evolution models have been developed to explain the
chemical composition of the solar vicinity and the [el/Fe] vs. [Fe/H] patterns (e.g. Henry et al. 2000;
Liang et al. 2001; Chiappini et al. 2003a, 2003b; Akerman et al. 2004, Fran\c cois 
et al. 2004). 
Other important constraints, which are connected  with the evolution
of the Galaxy disk, are the abundance gradients of the elements along the disk
of the Milky Way.
In general term, abundance gradients are a feature commonly observed in many
galaxies with their metallicities decreasing outward from the galactic centers.
The study of the gradients provides strong constraints to the mechanism of galaxy formation;
in fact, the star formation and the accretion history as function of the galactocentric distance
in the galactic disk influence strongly the formation and the development of the abundance gradients.
(see Matteucci\& Fran\c cois 1989, Boisser \& Prantzos 1999, Chiappini et al. 2001).
Many models have been already computed to explain the behaviours of abundances and abundance ratios as functions
of galactocentric radius (e.g  Hou et al. 2000; Chang et al. 1999;
 Chiappini et al. 2003b; Alib\'es et al. 2001) but they restrict
 their predictions only to a small number of chemical elements and do not consider very heavy elements.

We base our work on the chemical evolution model described in Chiappini et al.
(2001), which is able to well reproduce the gradients in the Milky Way for N, O, S and Fe.

In this work we calculate the behaviour of the largest  number of heavy elements
(O, Mg, Si, S, Ca, Sc, Ti, Co, V, Fe, Ni, Zn, Cu, Mn, Cr, Ba, La and Eu) ever considered in 
this kind of models.
In this way we are also able to test the the recent  nucleosynthesis 
prescriptions described in Fran\c cois et al. (2004) for the $\alpha$ and iron peak elements
 and in Cescutti et al. (2005) for the Eu and Ba, whereas the prescriptions for Lanthanum are
 newly calculated in this paper following the same approach adopted for Barium in the paper 
by Cescutti et al. (2005).

Chemical evolution models adopting the above nucleosynthesis prescriptions have already been shown
to reproduce the evolution of the abundances in the solar neighborhood. Here we extend our predictions
to the whole disk to check if these models can also reproduce the abundance gradients.
We compare our model predictions  with new  observational data collected 
by Andrievsky et al.(2002abc,2004) and Luck et al.(2003)
(hereafter 4AL). They measured the abundances of all the selected elements (except Ba)
in a  sample of 130 galactic Cepheids found in the galactocentric distance range from 5 to 17 kpc.
In addition to the data by 4AL we also compare our theoretical predictions
with abundance measurements in giants and open cluster located at even larger 
galactocentric distances.

The paper is organized as follows: in Section 2 we present the observational 
data, in Section 3 the chemical evolution model is presented and in Section 
4 the adopted nucleosynthesis prescriptions  are described.
In Section 5  we present the results  and in Section 6 some conclusions are drawn.

\section{Observational data}\label{data}

In this work we use  the data by 4AL for all the studied elemental gradients.
 These accurate data have been derived for a large sample of galactic Cepheids.
Cepheids variables have a distinct role in the determination of radial abundance
gradients for a number of reasons. First, they are usually bright enough that they can 
be observed at large distances, providing accurate abundances; second, their distances are generally 
well determined, as these objects are often used as distance calibrators (see Feast \& Walker 1987);
 third, their ages are also well determined, on the basis of relations involving their periods, 
luminosities, masses and ages (see Bono et al. 2005). 
They generally have ages close to a few hundred millions years.
So we can safely assume that they are representative of the present day gradients.
The 4AL sample contains abundance measurements for 130 Cepheid stars located 
between  5 and 17 kpc  from the Galactic center, for all the elements we  want to study
but Ba. 
The advantage of this data is that it constitutes an homogeneous sample for a large number
of stars and measured elements. Therefore, the abundance gradients can be better traced
with better statistics. Moreover, only for Cepheids is possible to obtain
abundances for so many elements, as well as a good estimate of the distance, necessary to compute
the gradients.
4AL obtained multiphase observations for the great majority of stars.
For the distant Cepheids they used 3-4 spectra in order to 
derive the abundances, while for nearby stars 2-3 spectra were used.
Besides this data set, we also adopt the data by Yong et al.(2006), who
computed the chemical abundances of Fe, Mg, Si, Ca, Ti, La and Eu for 30 Cepheids stars. 
Among these 30 stars, we choose only the 20 which are not in common with the sample
of 4AL. We apply the off-set found by Yong et al.(2006) with respect to 
the work of 4AL, in order to homogenize the two samples.

In order to compare the results on Cepheids with another class of young objects,
we also use  the data of Daflon \& Cuhna  (2004). Their database contains abundances
of C, N, O, Mg, Al, Si, S for 69 late O- to early B-type star members of 25 OB associations,
open clusters and HII regions. They determine the mean abundances of the different
cluster or association of young objects. In fact, these objects have all ages under 50 Myr.
Therefore, we assume that they also represent the present day gradients.

With the purpose of extending the comparison between our model and the 
observational data toward the outer disk,
we also include the datasets of Carraro et al.(2004)
 and Yong et al.(2005), these authors observed distant open clusters up to 22 kpc.
In this case we also show the average values of individual stars belonging to a cluster.
These stars are red giants with an estimated age ranging  from  2 Gyr to 5 Gyr.
We compare these data with the results of our model at the sun formation epoch,
 i.e. 4.5 Gyr before the present time.
Yong et al. (2005) measured the surface abundances of O, Mg, Si, Ca, Ti, Mn, Co,
Ni, Fe La, Eu and Ba for 5 clusters, whereas Carraro et al. (2004) computed the surface abundances 
of O, Mg, Si,Ca, Ti, Ni, Fe in 2 clusters. We underline that one of the clusters of Carraro et al. (2004), 
Berkeley 29, is in common with the sample of Yong et al.(2005) and we show  both measurements.
The galactocentric distance of this  object is 22 kpc and hence is the most distant open cluster
 ever observed.

Finally, we show the abundances of three field red giants,  which have been identified
 in the direction of the southern warp of the Galaxy by Carney et al.(2005).
In their work, they measure the abundances of O, Mg, Si, Ca, Ti, Mn, Co, Ni, Fe, La, Eu and Ba 
for the three red giants. The galactocentric distance of these object ranges from 10 kpc to 15 kpc.
The age of these three stars is unknown but it is likely that it is similar to the age of the
 red giants measured in the old open clusters. Therefore, we may compare them with the abundances at the 
solar system formation time.

\section{The chemical evolution model for the Milky Way}

In our model, the Galaxy is assumed to have formed by means of two main infall episodes:
 the first forms the halo and the thick disk, the second the thin disk.
 The timescale for the formation of the halo-thick disk is 0.8 Gyr. The timescale 
for the thin disk is much longer, 7Gyr in the solar vicinity, implying that the infalling gas
 forming the thin disk comes mainly from the  intergalactic medium and not only from the halo
 (Chiappini et al. 1997). 
Moreover, the formation of the thin disk is assumed to be a function of the galactocentric distance,
 leading to an inside out scenario for the Galaxy  disk build-up (Matteucci \& Fran\c cois 1989).
The galactic thin disk is approximated by several independent rings, 2 kpc wide, without 
exchange of matter between them.

 The main characteristic  of the two-infall model is an almost independent evolution
 between the halo and the thin disk
(see also Pagel \& Tautvaisienne 1995).
 A threshold gas density of $7M_{\odot}pc^{-2}$ in the star formation process (Kennicutt 1989, 1998,
 Martin \& Kennicutt 2001) is also adopted for the disk.

 The model well reproduces already an extended set of observational constraints in particular for the solar
neighborhood. Some of the most important observational constraints
are represented by the various relations between the abundances of metals (C,N,O,$\alpha$-elements,
iron peak elements) as functions of the [Fe/H] abundance  (see Chiappini et al. 2003a, b and
Fran\c cois et al. 2004) and by the G-dwarf metallicity distribution.
 It is worth mentioning here that, although this model is probably not unique,
however it reproduces the majority of the observed features of the Milky Way.
Many of the assumptions of the model  are shared by other authors (e.g. Prantzos \& Boissier
2000, Alib\'es  et al. 2001, Chang et al. 1999).

Chiappini et al.(2001) have shown that the chemical evolution of the halo can
have an impact in the abundance gradients in the outer parts of the disk.
They  analyzed the influence of the halo surface mass density on
the formation of the abundance gradients  of  O, S, Fe and N at large galactocentric distances.
In their model A, the halo surface mass density is assumed constant and equal to  $17M_{\odot}pc^{-2}$
for $R\le8 kpc$ and decreases as $R^{-1}$ outward. A threshold in the gas density 
is assumed also for the halo and set to $4M_{\odot}pc^{-2}$.
Then model B has a constant surface mass density  equal to  $17M_{\odot}pc^{-2}$ 
 for all the galactocentric distances and the threshold in the halo phase is the same as in 
 model A. 
In their model C the halo surface mass density is assumed as in model A 
but it does not have a threshold in the halo phase.
In their model D, both the halo surface mass density and the threshold is as in model A
but the time scale for the halo formation at galactocentric distances larger than 10 kpc
is set to 2 Gyrs and to 0.8 Gyr for distances smaller that 10 kpc. In all the other models,
the halo timescale is constant for all the galactocentric distances and equal to 0.8 Gyr.
 Here we will show our model predictions mainly for model B.
 In fact, the differences among model A,B,C and D arise primarily in the predicted steepness
 of the gradients for the outermost disc regions of the galactic disc. 
In this zone, the model B predicts the flattest gradients among the models 
of Chiappini et al. (2001), and provides the best fit according to observed 
flatness in the recent data by 4AL and in the distant open clusters.
Model A is also a good agreement with the abundance gradients traced by Cepheids up to
$\sim$ 12 kpc, whereas for larger galactocentric distances this model tends to be systematically
below the observations. We will show the predictions of this model only for the $\alpha$-elements.
 The model C shows a trend similar to the model A, whereas the model D tends to be below the observations 
already for galactocentric distances greater than $10kpc$, so we chose to not show their 
predictions.
We do not give here a detailed description of the model that can be 
found in Chiappini et al. (2001); nevertheless, to better understand
how the gradients form, it is fundamental to know how we model
the built up of the disk and the halo, so the rate of  mass accretion  A(r,t),
which is a function of time and galactocentric distance:

\begin{equation}
A(r,t)=a(r)e^{-t/\tau_{H}}+b(r)e^{(t-t_{max})/\tau_{D}(r)}.
\end{equation}

In this equation, $t_{max}=1Gyr$ is the time for the maximum infall rate on the thin disk,
 $\tau_{H}=0.8Gyr$ is the time scale
for the formation of the halo thick-disk and $\tau_{D}$ is the timescale of the thin disk,
which is a function of the galactocentric distance:
\begin{equation}
\tau_{D}=1.033r(kpc)-1.267Gyr.
\end{equation}
The coefficients $a(r)$ and $b(r)$ are constrained to reproduce the present day
 total surface mass density as a function of galactocentric distance.
In particular, $b(r)$ is assumed to be different from zero only for $t>t_{max}$,
 where $t_{max}$ is the time of maximum infall on the thin disk
 (see Chiappini et al. 2003a, for details).
Another important ingredient of the model is the adopted
law for the SFR, which is the following:
\begin{equation}
\psi(r,t)=\nu\left(\frac{\Sigma(r,t)}{\Sigma(r_{\odot},t)}\right)^{2(k-1)}
\left(\frac{\Sigma(r,t_{Gal})}{\Sigma(r,t)}\right)^{k-1}G^{k}_{gas}(r,t).
\end{equation}
$\nu$ is the efficiency of the star formation process and is set to be $2Gyr^{-1}$
for the Galactic halo ($t<1Gyr$) and $1Gyr^{-1}$ for the disk ($t\ge1Gyr$).
$\Sigma(r,t)$ is the total
surface mass density, $\Sigma(r_{\odot},t)$ the total surface mass density at the 
solar position, $G_{gas}(r,t)$ the surface density normalized to the present time
total surface mass density in the disk $\Sigma_{D}(r,t_{Gal})$, where $t_{Gal}=13.7Gyr$ is the age 
assumed for the Milky Way and $r_{\odot}=8kpc$ the solar galactocentric distance 
(Reid 1993). The exponent of the surface gas density, $k$, is set equal to 1.5.
With these values for the parameters the observational constraints, in particular in the solar vicinity,
are well fitted.
We recall that below a critical threshold for the gas surface density 
($7M_{\odot}pc^{-2}$ for the thin disk and $4M_{\odot}pc^{-2}$ for the halo phase)
 we assume no star formation.

\begin{figure}
\begin{center}
\includegraphics[width=0.35\textwidth]{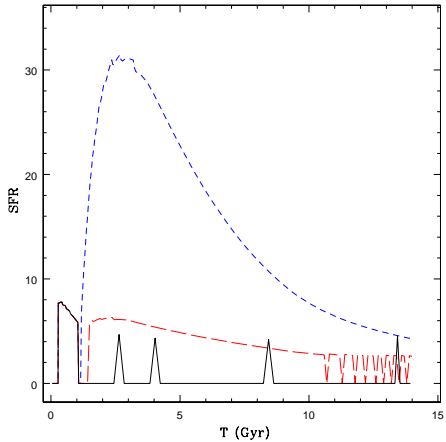}
\caption{ The SFR expressed  in $M_{\odot}pc^{-2}Gyr^{-1}$, as predicted by the two infall model for 
different Galactocentric distances: 4 kpc (short dashed line), 8 kpc (long dashed line) and 12 kpc 
(solid line).
The SFR in the halo phase  (indicated by the continous line) up to 0.8 Gyr,
 is the same for all the galactocentric distances, 
whereas in the disk the SFR changes according to the different infall rates. 
Note that at 4 kpc distance the SFR in the disk is much higher 
than at larger galactocentric distances.
The gap in the SFR at the end of the halo-thick disk phase is evident in the solar neighborhood.
The oscillations are due to to the threshold density.}

\label{SFR}
\end{center}
\end{figure}

In Fig. (\ref{SFR}) we  show the predicted star formation rate for three different 
galactocentric distances: 4, 8 and 12 kpc;
the SFR is the same for every galactocentric distance during the halo phase,
 due to the fact that the assumed halo mass density in the selected model B
 is not a function of galactocentric distance; the critical threshold of the gas surface 
density  naturally produces  a bursting star formation history in the outer part of the disk,
  whereas at the solar neighborhood, it happens only toward the end of the evolution.
We note that at the solar galactocentric distance, which is assumed to be 8 kpc,
 the threshold also produces a hiatus between the halo phase and the thin disk phase.

\section{Nucleosynthesis Prescriptions}{\label{NP}}

\subsection{$\alpha$ and Iron peak elements}
For the nucleosynthesis prescriptions of the Fe and the others elements (namely O, S, Si, Ca, Mg, Sc, Ti,
 V, Cr, Zn, Cu, Ni, Co and Mn ), we adopted those suggested in Fran\c cois et al. (2004). 
They compared theoretical predictions about the [el/Fe] vs. [Fe/H] trends in 
the solar neighborhood for the above mentioned elements and they selected the best sets
of yields required to best fit the data.
In particular for the yields of SNe II they found that the Woosley \& Weaver (1995) ones
 provide the best fit. In fact, no modifications are required
for the yields of Ca, Fe, Zn and Ni as computed for solar composition. For Oxygen the best results 
are given by the Woosley \& Weaver (1995) yields computed as functions of the metallicity.
For the other elements, variations in the predicted yields are required to best fit the data
(see Fran\c cois et al. 2004 for details).
For what concerns the yields from type SNeIa, revisions in the theoretical yields by Iwamoto et al.(1997) 
are requested for Mg, Ti, Sc, K, Co, Ni and Zn to best fit the data.
The prescription for single low-intermediate mass stars are by van den Hoek \& Groenewegen (1997),
for the case of the mass loss parameter which varies with metallicity (see Chiappini et al. 2003a, model5).

\subsection{S and R process}{\label{NP_BaS}}
For the nucleosynthesis of s-process we have adopted
 the yields of Busso et al. (2001) in the mass
range 1.5-3$M_{\odot}$ for Lanthanum and Barium. 

The theoretical results by Busso et al. (2001) suggest 
negligible Europium production  in the s-process and therefore
 we neglected this component in our work.
We have extended the theoretical results of Busso 
et al. (2001)  in the mass range $1.5-1M_{\odot}$,
by simply scaling with the mass the values obtained for stars of $1.5 M_{\odot}$.
We have extended the prescription in order to better fit the data with a [Fe/H]
higher than solar. This hypothesis does not change the results of the model at [Fe/H]$<$0.

For the nucleosynthesis prescriptions of r-process elements
we used the model 1 by Cescutti et al. (2005) for both Ba and Eu.
These empirical yields have been chosen in order to reproduce 
the surface abundances for Ba and Eu  of low metallicity stars 
as well as the Ba and Eu solar abundances. 
Cescutti et al.(2005) have assumed that Ba is also produced as an r-process element
in massive stars (12 - 30 $M_{\odot}$), whereas  Eu is considered 
to be a purely r-process element produced in the  same range of masses.

For La we give new prescriptions  following the same method as for Ba:
 we assume an r-process contribution in massive stars (12 - 30 $M_{\odot}$), besides
the s-process contribution from low mass stars.
The yields of this r-process contribution are summarized in Table \ref{rLa},
in which the mass fraction of newly produced La is given as function of the
mass.
The results for the solar neighborhood are shown in Sect. 5.

\begin{table*}

\caption{The stellar yields for La in massive stars (r-process)
in the case of primary origin.} \label{rLa}

\centering

\begin{minipage}{90mm}

\begin{tabular}{|c|c|c|}
\hline

$M_{star}$  & $ X_{La}^{new}$\\

\hline\hline

12.   & 9.00$\cdot10^{-8}$ \\ 
15.   & 3.00$\cdot10^{-9}$ \\   
30.   & 1.00$\cdot10^{-10}$ \\

\hline\hline

\end{tabular}

\end{minipage}

\end{table*}

\begin{table*}

\caption{The mean and the standard deviations for the abundance of [La/Fe] for the stars 
inside each  bins along the [Fe/H] axis.} \label{meanLa}

\begin{tabular}{|c|c|c|c|c|}
\hline

bin center [Fe/H]& bin dim.[Fe/H]  & mean [La/Fe] & SD [La/Fe] &  N. of data in the bin \\
\hline\hline

 -2.97 & 1.20 & -0.13&  0.48& 29\\
 -2.17 & 0.40 &  0.04&  0.26& 15\\
 -1.78 & 0.40 &  0.06&  0.17& 7 \\
 -1.38 & 0.40 &  0.18&  0.17& 5 \\
 -0.99 & 0.40 & -0.04&  0.29& 4 \\
 -0.59 & 0.40 &  0.19&  0.32& 5 \\
 -0.19 & 0.40 &  0.09&  0.14& 13\\
  0.20 & 0.40 & -0.08&  0.09& 7 \\

\hline \hline
       
\end{tabular}

\end{table*}

\section{Results for the solar vicinity}

\begin{figure}
\begin{center}
\includegraphics[width=0.35\textwidth]{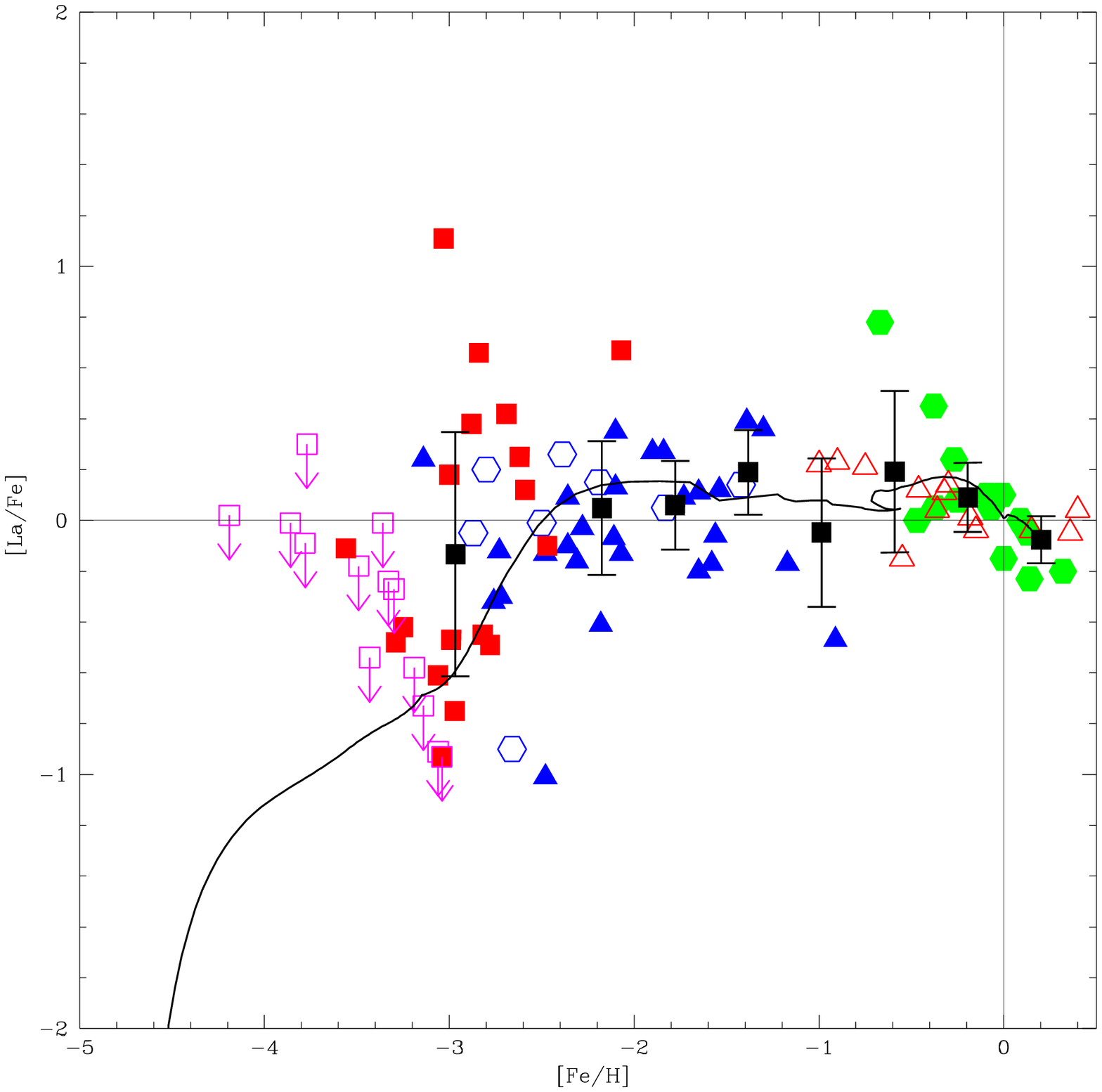}
\caption{In this Fig. is  plotted the [La/Fe] versus [Fe/H]. The data are taken
 from Fran\c cois et al.(2006), (filled red squares, whereas the pink open
 squares are only upper limits), Cowan et al.(2005)
(blue open hexagons), Venn et al. (2004) (blue solid triangles), Pompeia et al.
(2003) (green filled hexagons) and McWilliam \& Rich (1994)  (open red triangles).
 The black squares are  the mean values of the data bins described in the Table \ref{meanLa}.
 As error bars we consider the standard deviation (see Table \ref{meanLa}).
The solid line is the results of our model for La (see Table \ref{rLa}), normalized 
to the solar abundance as measured by Asplund et al.(2005)}
\label{Laresult}
\end{center}
\end{figure}

We present here the new results for the solar neighborhood.
To better investigate the trends of the data we divide the 
[Fe/H] axis in several bins 
 and we compute the mean and the standard deviations from the mean of the ratios
[La/Fe] for all the data inside each bin. These results are shown in Table \ref{meanLa},
where we also summarize the center and the dimension of each bin and the number of data
contained in each of them.
In Fig.\ref{Laresult} we show  the predictions of the chemical 
evolution  model  for La in the solar neighborhood using our prescriptions for the yields in
 massive stars and the prescriptions of Busso et al. (1999) for low mass stars, 
as described in the previous section.
These results are new and the model well reproduces the trend 
of the stellar abundances at different [Fe/H] and the solar abundance 
of Lanthanum. In fact, we obtain a La mass fraction of $1.35 \cdot 10^{-9}$
substantially equal to the solar value of Asplund et al.(2005) of $1.38 \cdot 10^{-9}$, 
as shown in Table \ref{t:sun} where the predicted and observed solar abundances
are compared for all the elements studied here.  

We point out that the yields adopted for $\alpha$ and iron peak elements in Fran\c cois et al.(2004)
and for Ba and Eu in Cescutti et al.(2005) have already been shown to well fit observational data 
in the solar neighborhood. In the next sections we check whether with the same  
nucleosynthesis prescriptions  our model can explain the data in the other parts of the galactic disk.

\begin{table*}
\caption{Element abundances by Asplund et al.(2005) 
in the present-day solar photosphere and in meteorites (C1 chondrites) 
 compared to the results of our model at the solar formation epoch.} \label{t:sun}
\begin{minipage}{90mm}

\smallskip
\begin{tabular}{llccc|llccc}
\hline\hline

 & Elem.  &  Photosphere  &   Meteorites  & Model & & Elem.  &  Photosphere  &   Meteorites  & Model \\

\hline

8  & O   &  $8.66 \pm 0.05$   &  $8.39 \pm 0.02$ &  8.67 &25 & Mn  &  $5.39 \pm 0.03$   &  $5.47 \pm 0.03$ &  5.44     \\
12 & Mg  &  $7.53 \pm 0.09$   &  $7.53 \pm 0.03$ &  7.57 &26 & Fe  &  $7.45 \pm 0.05$   &  $7.45 \pm 0.03$ &  7.41    \\
14 & Si  &  $7.51 \pm 0.04$   &  $7.51 \pm 0.02$ &  7.58 &27 & Co  &  $4.92 \pm 0.08$   &  $4.86 \pm 0.03$ &  4.88    \\
16 & S   &  $7.14 \pm 0.05$   &  $7.16 \pm 0.04$ &  7.20 &28 & Ni  &  $6.23 \pm 0.04$   &  $6.19 \pm 0.03$ &  6.23   \\
20 & Ca  &  $6.31 \pm 0.04$   &  $6.29 \pm 0.03$ &  6.25 &29 & Cu  &  $4.21 \pm 0.04$   &  $4.23 \pm 0.06$ &  4.13   \\
21 & Sc  &  $3.05 \pm 0.08$   &  $3.04 \pm 0.04$ &  3.05 &30 & Zn  &  $4.60 \pm 0.03$   &  $4.61 \pm 0.04$ &  4.53    \\
22 & Ti  &  $4.90 \pm 0.06$   &  $4.89 \pm 0.03$ &  4.90 &56 & Ba  &  $2.17 \pm 0.07$   &  $2.16 \pm 0.03$ &  2.19    \\
23 & V   &  $4.00 \pm 0.02$   &  $3.97 \pm 0.03$ &  3.59 &57 & La  &  $1.13 \pm 0.05$   &  $1.15 \pm 0.06$ &  1.11    \\
24 & Cr  &  $5.64 \pm 0.10$   &  $5.63 \pm 0.05$ &  5.59 &63 & Eu  &  $0.52 \pm 0.06$   &  $0.49 \pm 0.04$ &  0.56   \\

\hline\hline
\end{tabular}

\end{minipage}

\end{table*}

\section{Abundance gradients compared with the 4AL data}

We run the model described in Sect.4 and we predict the variation of the 
abundances of 
the studied elements along the galactic disk in the galactocentric range 
5 - 17 kpc, at the
present time.
We then compare the abundances predicted by our model at the present time 
for all the elements with the observational data.
To better understand the trend of the data, we choose to divide the data in 6 bins as functions 
of the galactocentric distance. In each bin we compute the mean value and the standard deviations for
all the elements. The results are shown in Table \ref{meanAbunda}:
in the first column we show the galactocentric distance range chosen for each bin,
 in the second column the mean galactocentric distance for the stars inside the considered bin,
 in the  other columns the mean and the standard deviation 
computed for the abundances of every chemical elements, inside the considered bin.
We note that for some stars it has not been possible to measure all the abundances.
We plot the results of our model at the present day normalized both 
to the solar observed abundances by Asplund et al.(2005) and to the mean 
value of the abundance data by 4AL at the solar distance.
For some elements, in fact, there is a discrepancy between
 the predicted abundances at the present day by our model and the mean abundances  
 of the observed Cepheids at the solar distance.

\begin{table*}

\caption{The mean value and the standard deviation inside each bin for for O, Mg, Si and S.} 
\label{meanAbunda}

\begin{minipage}{90mm}

\begin{tabular}{|c|c|c|c|c|c|c|c|c|c|c|c||c|c|c|c|c|}
\hline

 galactocentric    & mean GC      &   mean  &   SD    &  mean   &  SD    & mean   &  SD    & mean &  SD       \\
 distance range    & distance(kpc)&  [O/H]  &  [O/H]  &  [Mg/H] & [Mg/H] & [Si/H] & [Si/H] & [S/H]& [S/H]    \\  
\hline	  
         $<$6.5kpc &   5.76       &   0.16  &    0.17 &  -0.19  &  0.17  &  0.21  &  0.13  &  0.37 & 0.19     \\ 
 6.5$<$--$<$7.5kpc &   7.10       &  -0.08  &    0.13 &  -0.19  &  0.10  &  0.07  &  0.06  &  0.17 & 0.08     \\
 7.5$<$--$<$8.5kpc &   8.00       &  -0.06  &    0.13 &  -0.19  &  0.13  &  0.06  &  0.06  &  0.09 & 0.10     \\ 
 8.5$<$--$<$9.5kpc &   8.96       &  -0.12  &    0.16 &  -0.21  &  0.11  &  0.04  &  0.04  &  0.08 & 0.17     \\ 
 9.5$<$--$<$11 kpc &  10.09       &  -0.16  &    0.19 &  -0.22  &  0.18  & -0.07  &  0.07  & -0.11 & 0.20     \\
         $>$11 kpc &  12.33       &  -0.19  &    0.21 &  -0.32  &  0.13  & -0.16  &  0.08  & -0.23 & 0.15     \\
\hline\hline

 galactocentric    & mean GC      & mean   & SD   & mean   &  SD     & mean   &  SD     &   mean   &  SD      \\
 distance range    & distance(kpc)& [Ca/H] &[Ca/H]& [Sc/H] & [Sc/H]  & [Ti/H] & [Ti/H]  &   [V/H]  & [V/H]    \\  
 \hline	  	                   	  	 
 $<$6.5kpc &   5.76       &  0.11  & 0.19&   0.15  &   0.18  &  0.19  &   0.13  &    0.14 &   0.14     \\ 
 6.5$<$--$<$7.5kpc &   7.10       &  0.00  & 0.10&  -0.05  &   0.18  &  0.05  &   0.08  &    0.04 &   0.05     \\
 7.5$<$--$<$8.5kpc &   8.00       & -0.04  & 0.07&  -0.06  &   0.13  &  0.05  &   0.07  &    0.03 &   0.09     \\ 
 8.5$<$--$<$9.5kpc &   8.96       & -0.04  & 0.11&  -0.09  &   0.13  &  0.04  &   0.06  &   -0.01 &   0.09     \\ 
 9.5$<$--$<$11 kpc &  10.09       & -0.13  & 0.09&  -0.12  &   0.09  & -0.05  &   0.14  &   -0.08 &   0.15     \\
 $>$11 kpc &  12.33       & -0.19  & 0.11&  -0.21  &   0.11  & -0.15  &   0.08  &   -0.21 &   0.11     \\
 \hline\hline

 galactocentric    & mean GC      & mean   & SD     &  mean   & SD    & mean   & SD     & mean   & SD     \\
 distance range    & distance(kpc)& [Cr/H]  & [Cr/H]&  [Mn/H] & [Mn/H]& [Fe/H] & [Fe/H] & [Co/H] & [Co/H] \\
 \hline	  	                                     	  	                    
 $<$6.5kpc &   5.76       &  0.11  &   0.11 &  0.09   &  0.14 &  0.17  &   0.13 &  0.06  &   0.13 \\
 6.5$<$--$<$7.5kpc &   7.10       &  0.05  &   0.08 &  0.05   &  0.12 &  0.05  &   0.07 & -0.10  &   0.07 \\
 7.5$<$--$<$8.5kpc &   8.00       &  0.04  &   0.11 &  0.01   &  0.11 &  0.01  &   0.06 & -0.10  &   0.11 \\
 8.5$<$--$<$9.5kpc &   8.96       &  0.03  &   0.12 &  0.00   &  0.09 & -0.01  &   0.08 & -0.06  &   0.11 \\
 9.5$<$--$<$11 kpc &  10.09       & -0.06  &   0.12 & -0.18   &  0.13 & -0.09  &   0.09 & -0.17  &   0.18 \\
 $>$11 kpc &  12.33       & -0.20  &   0.10 & -0.31   &  0.18 & -0.22  &   0.09 & -0.14  &   0.17 \\
 
 \hline

 \end{tabular}
 
 \begin{tabular}{|c|c|c|c|c|c|c|c|c|c|c|c|}
 
 \hline
 galactocentric & mean GC     &  mean  & SD    & mean  & SD    & mean    & SD    & mean   & SD     & mean   & SD     \\
 distance range &distance(kpc)& [Ni/H] & [Ni/H]&[Cu/H] &[Cu/H] & [Zn/H]  & [Zn/H]& [La/H] & [La/H] & [Eu/H] & [Eu/H] \\
 \hline  	                                                   	  	                   
 $<$6.5kpc &  5.76 &  0.18 &  0.16 & 0.15  &  0.18 &  0.51   &  0.20 &  0.21  &  0.10  & 0.17 &   0.14 \\
 6.5$<$--$<$7.5kpc &  7.10 &  0.02 &  0.07 & 0.07  &  0.12 &  0.28   &  0.10 &  0.19  &  0.07  & 0.05 &   0.06 \\
 7.5$<$--$<$8.5kpc &  8.00 & -0.02 &  0.08 & 0.08  &  0.30 &  0.26   &  0.12 &  0.22  &  0.06  & 0.08 &   0.08 \\
 8.5$<$--$<$9.5kpc &  8.96 & -0.04 &  0.09 & 0.12  &  0.19 &  0.28   &  0.14 &  0.26  &  0.08  & 0.08 &   0.10 \\
 9.5$<$--$<$11 kpc & 10.09 & -0.14 &  0.10 &-0.35  &  0.29 &  0.16   &  0.09 &  0.23  &  0.09  & 0.04 &   0.13 \\
 $>$11 kpc & 12.33 & -0.23 &  0.12 &-0.09  &  0.17 &  0.10   &  0.11 &  0.12  &  0.11  & 0.00 &   0.14 \\
 
 \hline\hline

\end{tabular}

\end{minipage}

\end{table*}

\begin{center}
\begin{table*}

\caption{Model results for present time  gradients for each element. We show the gradients computed as 
a single slope, for all the range of galactocentric distance considered,  and as two slopes: 
from 4 to 14Kpc and from 16 to 22Kpc.}\label{grad:}
\begin{minipage}{90mm}

\smallskip
\begin{tabular}{l|ccccccccc}
\hline\hline

\hline

 							   &         Fe &     O &    Mg &   Si &     S &    Ca  &   Cu  &    Zn &   Ni \\
 
\hline

 $\frac{\Delta [el/H]}{\Delta R}(dex/Kpc)$from 4 to 22 Kpc &	-0.036	&-0.028	&-0.031	&-0.033	&-0.034	&-0.034	&-0.050	&-0.038	&-0.034	\\
 $\frac{\Delta [el/H]}{\Delta R}(dex/Kpc)$from 4 to 14 Kpc &   	-0.052	&-0.035	&-0.039	&-0.045	&-0.047	&-0.047	&-0.070	&-0.054	&-0.047 \\
 $\frac{\Delta [el/H]}{\Delta R}(dex/Kpc)$from 16 to 22 Kpc&  	-0.012	&-0.011	&-0.012	&-0.012	&-0.012	&-0.012	&-0.014	&-0.012	&-0.012	\\

\hline\hline
&								Sc      &Ti	& V	& Cr	& Mn	& Co	& Ba	& Eu	& La   \\
\hline
 $\frac{\Delta [el/H]}{\Delta R}(dex/Kpc)$from 4 to 22 Kpc &    -0.036	&-0.032	&-0.038	&-0.036	&-0.038	&-0.037	&-0.021	&-0.030	&-0.021\\
 $\frac{\Delta [el/H]}{\Delta R}(dex/Kpc)$from 4 to 14 Kpc &	-0.051	&-0.043	&-0.056	&-0.052	&-0.057	&-0.055	&-0.032	&-0.036	&-0.032\\
 $\frac{\Delta [el/H]}{\Delta R}(dex/Kpc)$from 16 to 22 Kpc&   -0.012	&-0.012	&-0.011	&-0.012	&-0.011	&-0.011	&-0.009	&-0.013	&-0.008 \\

\hline\hline
\end{tabular}

\end{minipage}

\end{table*}
\end{center}

\subsection{$\alpha$-elements (O-Mg-Si-S-Ca)}

\begin{figure}
\begin{center}
\includegraphics[width=0.49\textwidth]{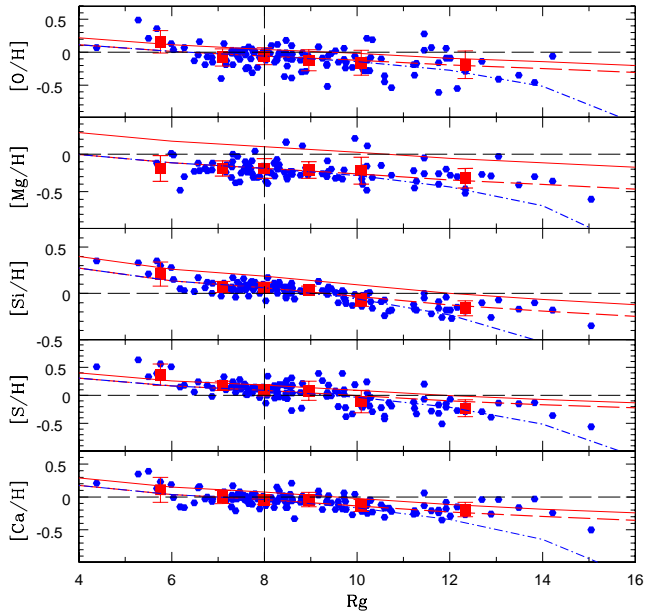}
\caption{We plot the abundances for O, Mg, Si, S and Ca as functions of the galactocentric distance.
The blue dots are the data by 4AL, the red squares are the mean values inside each bin only for 
the data by 4AL and the error bars are the standard deviations (see table \ref{meanAbunda}). 
The thin solid line is our model normalized at the observed solar abundances by Aslpund et al.(2005),
 whereas the thick dashed line is normalized at the mean value of the bin centered 
in 8 kpc (the galactocentric distance of the Sun). The dash-dotted
line is the results of the model with the prescriptions for the halo gas density of model A by Chiappini 
et al. (2001) normalized at the mean value of the bin centered in 8 kpc (cfr. Sect.3).}
\label{F1}
\end{center}
\end{figure}

We plot the results for these elements in Fig. \ref{F1}.
 We note that there is a discrepancy between our predictions normalized at the solar abundances by
Asplund et al. (2005) and the mean abundance of these elements for Cepheids at
the solar galactocentric distance. In fact, the predictions of our model for these elements at the 
present time at 8 kpc are supersolar, whereas the mean abundances of Cepheids 
for Mg, Ca and O are subsolar and this difference is particularly  
evident for Mg.
This means that either our model predicts a too steep increase of 
metallicity in the last 4.5 Gyr or, that the absolute abundances of Cepheids are 
underestimated. In fact, as 4AL underline in their first article, some uncertanties in 
the absolute abundances could exist, but the slope of the abundance 
distributions should be hardly affected.
 If these subsolar abundances were real, then one might think that they are the effect
 of some additional infall episode, occurring in the last 4.5 Gyr.
However, the goal of this work is to reproduce the trend of
the gradients, so here this problem can be neglected and we can compare
the data with the model results  normalized to
 the mean abundances of the Cepheids at 8 kpc.
These results well reproduce the trend of the data for all the five elements.
Moreover, in the case of Si we note that the data show a very little spread 
(as the small standard deviation values indicate)
and our model (the one normalized at the mean abundance at 8 kpc) perfectly 
lies over the mean value in each bin.
Finally, we note for S that the values predicted by the model for small galactocentric distances 
are inside the error bar of the data but a bit too low.
The trend for Ca is nicely followed by our model and the data for Ca show a very little spread.
 On the other hand, we note that that the trend of Mg shows a shallower 
 slope toward the galactic center than the other $\alpha$-elements.
 This is probably due to the lack of the Mg data for the stars located 
from 4 to 6.5 Kpc, which determine the steep slope for the other $\alpha$-elements.
The results of the model, if we use the prescriptions for the halo gas density of model A by Chiappini
 et al. (2001) well reproduce the data up to 10 kpc but the model B reproduces better the data
for larger galactocentric distances.
 For this reason in the next sections we will show only the result of the model B. 
 In Table 5  we show the slopes of the gradients for all the studied elements, as predicted by model B. The
 gradients become flatter towards the outermost disk regions, in agreement with the Cepheids data.
 It is worth noting that each element has a slightly different slope, due to the different production
 timescales and nucleosynthesis processes. In particular, $\alpha$-elements (O,Mg, Si, Ca etc..) have
 generally flatter slopes than the Fe-peak-elements. In addition, there are differences even among the
 $\alpha$-elements such as Si and Ca relative to Mg and O: the slightly steeper slope of Si and Ca is due
 to the fact that these elements are produced also by Type Ia SNe, whereas O and Mg are not. In general,
 elements produced on longer timescales have steeper gradients. 
This is confirmed by the observations not 
only of Cepheids but also of open clusters and HII regions. Finally, the predicted gradients 
for s- and r- process elements seem flatter  than all the others. The reason for this is that they 
are produced in very restricted stellar mass ranges producing an increase of their abundances at 
low metallicities until they reach a constant value for [Fe/H] $> -3.0$ (see Fig. \ref{Laresult}).
 We think that the variations between gradients can be statistically significant, in particular those derived from Cepheids:
all the Cepheids, in fact, have similar  
atmospheric parameters (atmospheric temperature, surface gravity), 
their relative abundances are much less affected than the absolute abundances
 by the effect
of using LTE models instead of recent NLTE, 3D models.
Therefore the gradients from Cepheids look rather firmly established.

\subsection{Iron peak elements (Sc-Ti-Co-V-Fe-Ni-Zn-Cu-Mn-Cr)}
\begin{figure}
\begin{center}
\includegraphics[width=0.49\textwidth]{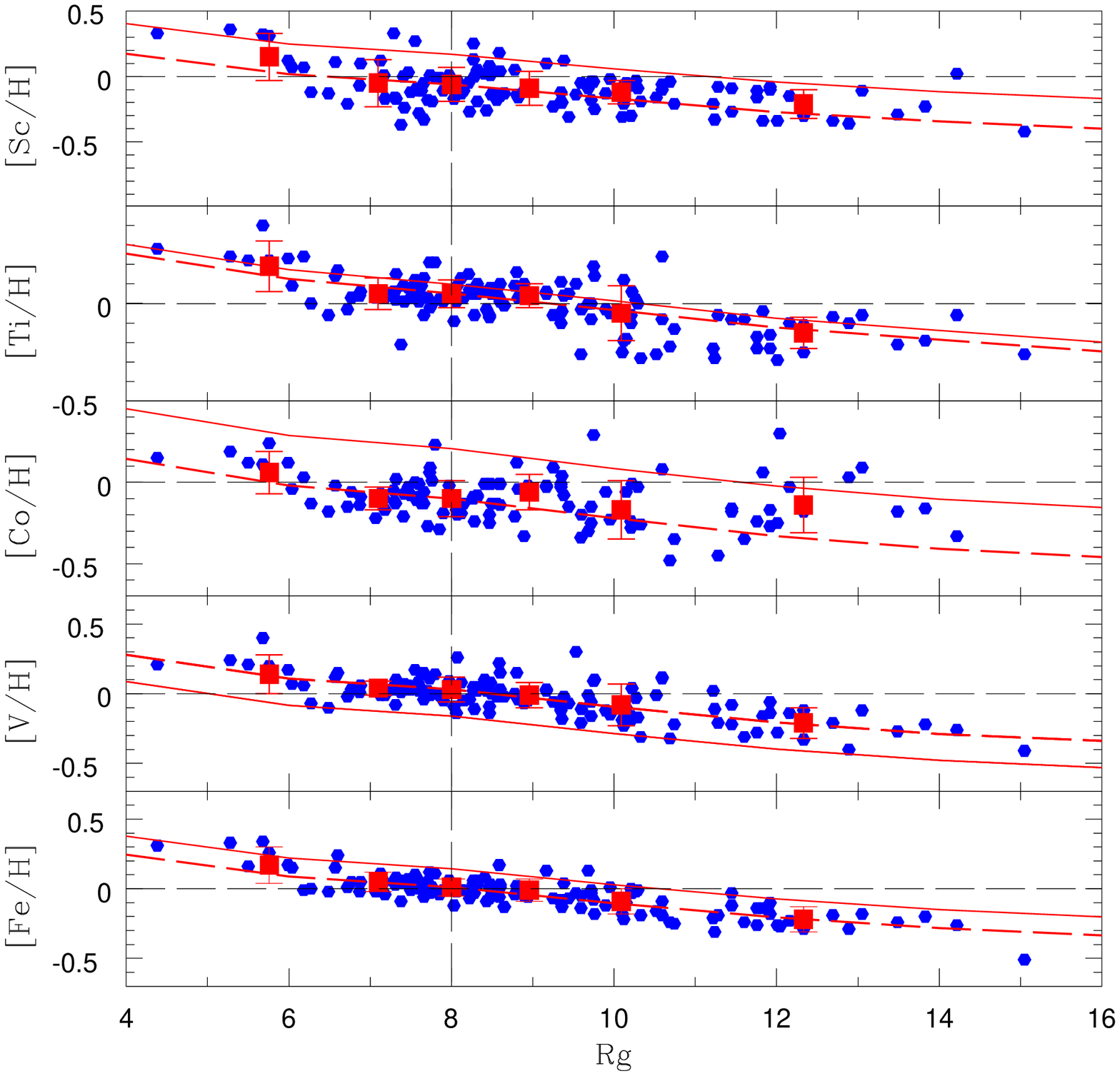}
\caption{Gradients for [Sc/H], [Ti/H], [Co/H], [V/H] and [Fe/H].
 The models and the symbols are the same as in Fig. \ref{F1}.}
\label{F2}
\end{center}
\end{figure}

\begin{figure}
\begin{center}
\includegraphics[width=0.49\textwidth]{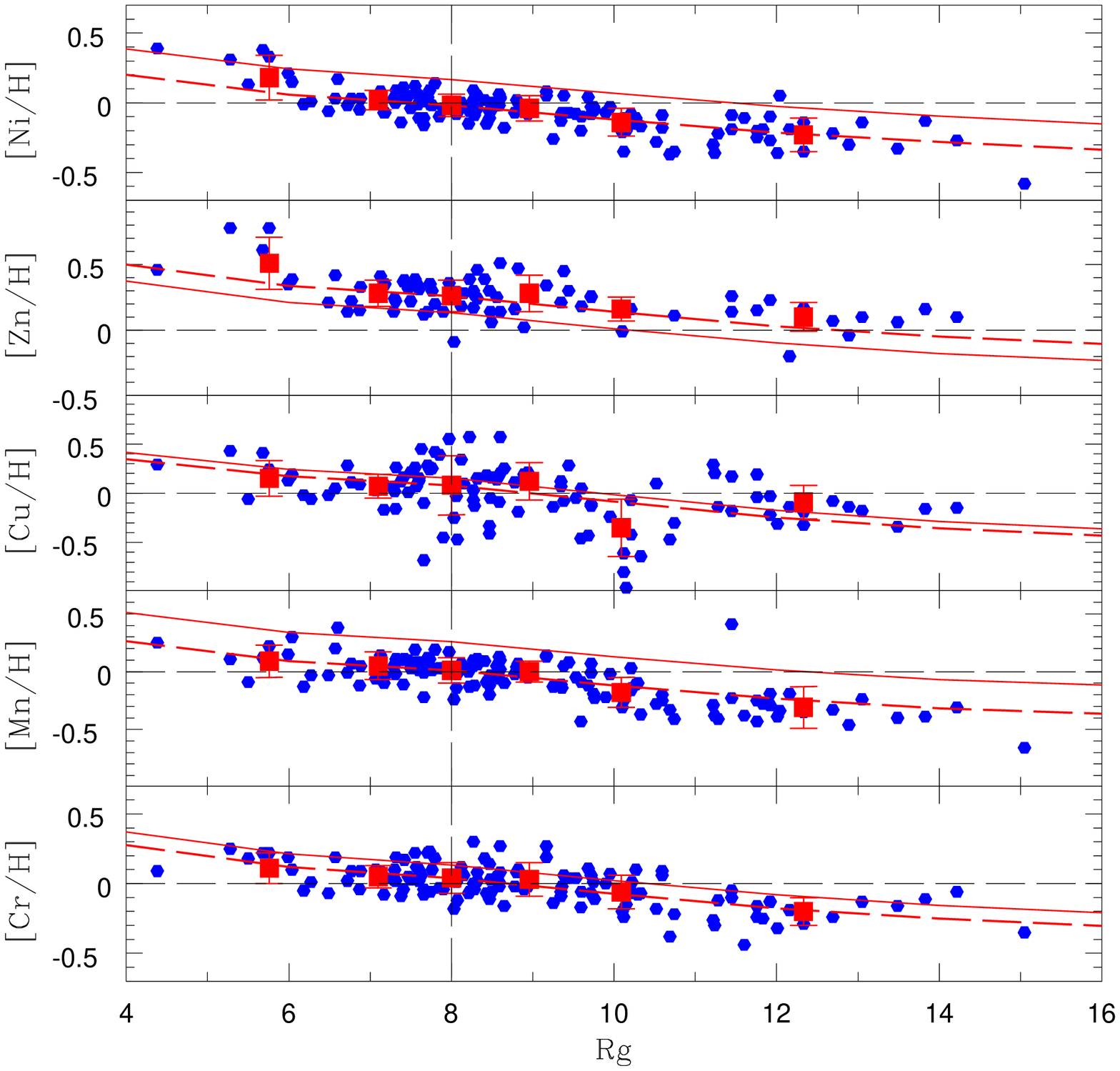}
\caption{Gradients for [Ni/H], [Zn/H], [Cu/H], [Mn/H] and [Cr/H].
 The models and the symbols are the same as in Fig. \ref{F1}.}
\label{F3}
\end{center}
\end{figure}

The ten elements of the so called iron peak are plotted in Figs. \ref{F2} and \ref{F3}.
The present time predictions of our model for
 iron peak abundances are super solar at 8 kpc, as for the $\alpha$-elements.
On the other hand, the mean values for iron peak elements 
in Cepheids in the bin at 8 kpc are in general solar, except Zn,
which is super solar and Sc and Co, which are sub solar.
Nevertheless the model gives a prediction for the trends of the gradients for these elements
 which is very good, in particular in the cases of V, Fe, Ni, Mn and Cr, as it is shown 
by the results of the model normalized to the mean value of the bin centered at 8kpc.
A problem is present for Co, for which a too low abundance is predicted by the model
 at galactocentric distances  $>$ 12 kpc.

\subsection{Neutron capture elements}

\begin{figure}
\begin{center}
\includegraphics[width=0.49\textwidth]{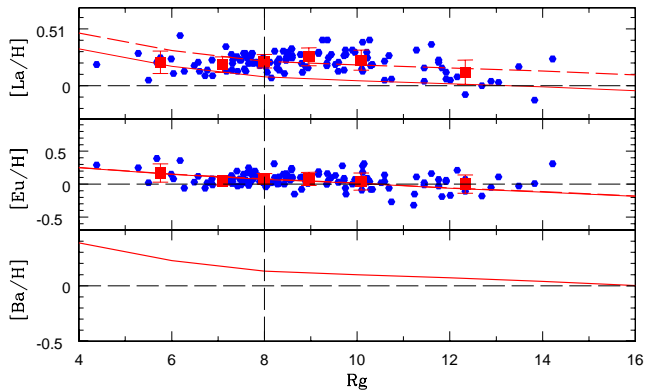}
\vspace{-3.2cm}
\caption{Gradients for [La/H], [Ba/H] and [Eu/H].
 The models and the symbols are the same as in Fig. \ref{F1}.
 Note that for Ba we show only the model predictions.}
\label{F4}
\end{center}
\end{figure}

It is well known that these elements present a large spread 
at low metallicities, which is not yet understood (see Cescutti et al. 2005
 and references therein).
 Since the Cepheids are young metal rich stars, this problem is not important.
 In fact, as shown in Fig. \ref{F4}, the  spread
in the data as a function of galactocentric distance is small. 
In the case of Eu our model well reproduces both the observed gradient and the mean value
for the Cepheid abundance at 8 kpc.
On the other hand, the mean value of the La abundance in the data by 4AL
at the solar galactocentric distance is about a factor 1.5 higher than the predicted abundance
by our model and the predictions for La show a slightly steeper gradients than the  data.
Always in  Fig. \ref{F4}, we show the predicted trend of the neutron capture
element Barium. For this element there are no data by 4AL; therefore, we just show our predictions
which have to be confirmed by future observations.

\section{Predicted abundance gradients compared with other sets of data}

We compare the results of our models with different sets of observational data, as
described in Sect. 2. We recall that only the data by Yong et al. (2006) refer to
Cepheids. However, these data and 4AL data are not homogeneous because of
 the different way in which the abundances are derived. As a consequence of this, we apply 
the offsets calculated by Yong et al. (2006),
on the basis of a representative sample of stars analyzed and measured by both authors,
to compare the two sets of data.
We underline that we compare the Cepheids  and the Daflon \& Cunha (2004) data for OB stars
with the model at the present time, normalized to the mean value at 8 kpc for the data
 by 4AL, whereas we compare observational data of red giants and open clusters
with the model at the solar formation time normalized at the  observed solar abundances 
by Asplund et al. (2005).

\subsection{$\alpha$-elements (O-Mg-Si-S-Ca)}

\begin{figure}
\begin{center}
\includegraphics[width=0.49\textwidth]{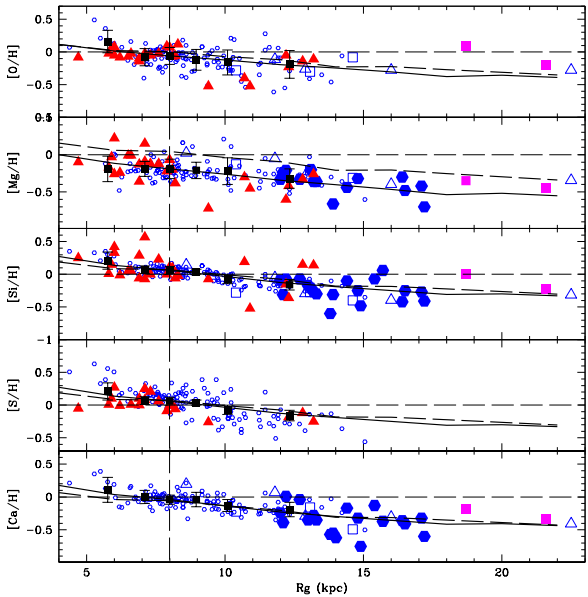}
\caption{The gradients for O, Mg, Si, S and Ca  compared with different sets of data.
The small  open circles are the data by 4AL, the black squares are the mean values inside each bin for 
the data by 4AL and the error bars are the standard deviations (see table \ref{meanAbunda}).
The red solid triangles are the data by Daflon \& Cunha (2004) (OB stars), the open blue 
squares are the data by Carney et al. (2005) (red giants),
the  blue solid hexagons are the data by Yong et al. (2006) (Cepheids), the blue open triangles are 
the data by Yong et al. (2005) (open clusters )and the magenta solid  squares are the
 data by Carraro et al. (2004) (open clusters).
Note that the most distant value for Carraro et al. (2004) and Yong et al. (2005)
refers to the same object: the open cluster Berkeley 29.
The thin solid line is our model at the present time normalized at the mean value 
of the bin centered in 8 kpc for Cepheids stars by 4AL; the dashed line represents
the predictions of our model at the epoch of the formation of the solar system normalized
to the observed solar abundances by Aspund et al. (2005).
This prediction should be compared with the data for red giant stars and open clusters 
(Carraro et al. 2004; Carney et al. 2005; Yong et al. 2005).}
\label{F5}
\end{center}
\end{figure}

In Fig. \ref{F5},  we show the comparison for for O, Mg, Si, S and Ca data with our model. 
Although the observations are from completely different types of 
astronomical objects (OB stars, red giants, open clusters and Cepheids),
 they  are substantially in agreement with each other and with our model. 
Nevertheless, the data by Yong et al. (2006) and the data by Daflon \& Cunha (2004) 
suffer a larger spread than the data by 4AL, in particular for Ca.
Finally the data by Carraro et al. (2004) on the open cluster Saurer 1 at 
the galactocentric distance of 18.7 kpc for all the considered elements, 
except Mg, are slightly above the prediction of our model.

\begin{figure}
\begin{center}
\includegraphics[width=0.49\textwidth]{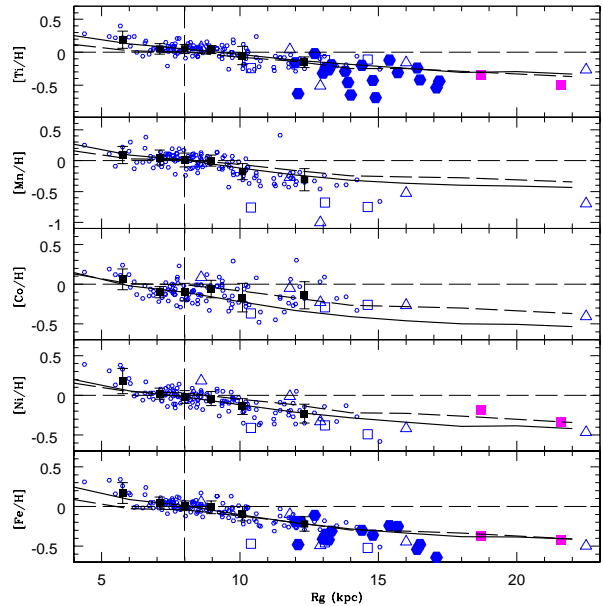}
\caption{Gradients for Ti, Mn, Co, Ni and Fe. The model and the symbols are the same
as  in Fig. \ref{F5}.}
\label{F6}
\end{center}
\end{figure}

\subsection{The iron peak elements (Ti-Mn-Co-Ni-Fe)}
We show the iron peak elements  in Fig. \ref{F6} . 
For these elements, it is possible to see some interesting features
in the observational data. The data by Yong et al.(2005)
seem to have a gradient in agreement with our model
if we take into account some possible offset in the data, as considered 
by Yong et al.(2006). In particular, the abundances of Mn in this data set are below
 our model predictions and the 4AL data.
We note that in the data by Yong et al.(2005), the open cluster Berkeley 31,
 which is at about 13 kpc, shows abundances lower than those predicted by our model and 
the set of data by 4AL for all the iron peak elements, with the exception of Co.
On the other hand, the set of data by Carney et al.(2005) shows an almost flat
trend and again lower abundances for the iron peak elements
  than the abundances of the data by 4AL and those predicted by the model.
This is probably due to the fact that the data are from
 old  and evolved objects, as giant stars are, with a not well estimated age.
The data by Yong et al. (2006), which consider the abundances for Ti and Fe,
are in agreement with our model and the data by 4AL, even if they seem
to present  slightly steeper gradients. 
Finally, the open cluster abundances as measured by Carraro et al.(2004) are in agreement
with our model, in particular for Fe, while for Ti and Ni the model fits both open 
cluster abundances inside the error bar, which is about 0.2 dex.

\begin{figure}
\begin{center}
\includegraphics[ width=0.49\textwidth]{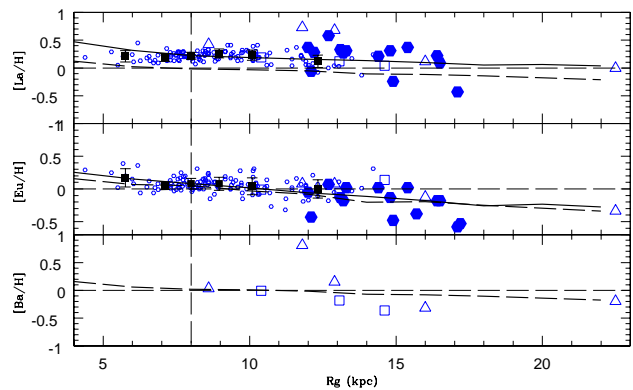}
\vspace{-3.2cm}
\caption{Gradients for La, Eu and Ba . The model and the symbols are the same
as  in Fig. \ref{F5}.}
\label{F7}
\end{center}
\end{figure}

\subsection{The neutron capture elements (La-Eu-Ba)}
We show the neutron capture  elements in Fig. \ref{F7}.
The data for Eu  are taken from the set of data by Yong et al.(2006) (Cepheids),
Yong et al. (2005) (open clusters) and Carney et al. (2005) (red giants),
 and they are nicely in agreement with our
model with the exception of a large spread in the data by Yong et al. (2006). 
Some problems arise for La. In fact, the trend of the gradients is similar for the sets of data
but the absolute values of the La abundances in the sets of data by Yong et al. (2006),
Yong et al. (2005) and Carney et al. (2005) are systematically lower than
the ones of 4AL  and, without the offset, it is impossible  to have a comparison.
Therefore, we apply an offset of +0.3 dex to all
observational data (with the exception of the 4AL data) to better 
show all the sets of data. 
We note that this can also be the conseguence of the different way of calculating the abundances,
as explained in Yong et al. (2006). It is worth noting that the most important results
are the slopes of the gradients rather than the absolute abundances.
With this offset applied to the data, the two open clusters (Berkeley 31 and NGC 2141),
 measured by Yong et al. (2005),
 still present  an abundance  of La larger than the one predicted by our model and the mean abundance 
of the data by 4AL; finally, the data of Yong et al.(2006) have again a large spread.
 Nevertheless, the comparison is pretty good and the abundance
of the most distant cluster is well fitted.
The results for Ba are similar to those for La. In fact, we have to
 apply an offset of +0.3 dex to the data by Carney et al. (2005),
for the same reasons explained above. The results are also
 quite good with the exceptions of the two open clusters
as before, which show a larger Ba abundance when compared 
to the results of our model.
We underline that for both these open clusters there is only a measured star 
and  it is possible that the stars chosen to be analyzed could be 
peculiar stars in terms of chemical abundances of s-process
elements and so they should not be considered  in the explanation of the gradients.

It is worth noting that the data in the outer parts of disk are
 still not enough to completely constrain our
models. Moreover, the existing samples show scatter. Two facts can affect the abundance 
gradients in the outer part: observational uncertainties in both, abundances and distance, and 
the fact that the outer parts could reflect a more complex chemical evolution. Moreover,
there are some suggestions that the open clusters and giants in the outer part of the disk
could have been accreted. However, despite these uncertanties, it is interesting to see 
that our chemical evolution model, where the halo density is assumed constant with radius out
to $\sim$ 20 kpc, predicts abundance substantially in agreement with those measured
 in the outer disk.

\section{Conclusions}

The aim of this work was to compare new observational data 
on the radial gradients for  17 chemical elements
with the predictions of our chemical evolution model for the Milky Way.
This model has been already tested on the properties of the solar vicinity and contains
a set of yields which best fit the abundances and abundance ratios in the solar vicinity,
 as shown in Fran\c cois et al. (2004).

The bulk of observational data comes from the abundances derived in 
a large number of Cepheids observed by
 4AL. For the first time, it is possible to verify the predictions for many heavy
elements with a statistical validity.

The comparison between model predictions and observational data showed that our model well reproduces
the gradients of almost all the elements that we analyzed. 
Since abundance gradients can impose strong constraints  both on the mechanism
of galaxy formation, in particular of the galactic disk, and the nucleosynthesis 
prescriptions, we can conclude that:

\begin{itemize}

\item The model for the Milky Way disk formation, assuming an inside-out building-up of the disk,
 as suggested originally by Matteucci \& Fran\c cois (1989), can be considered successful;
 in fact, for almost all the considered elements, we find a good fit
to the observational data ranging from 5 to 17 kpc. In particular, the model assuming
a constant total surface mass density for the halo best fits the data of 
Cepheids. In fact, at large galactocentric distances the halo mass distribution influences 
the abundance gradients (see Chiappini et al. 2001).

\item 
In our chemical evolution model we adopt a threshold in the gas density for star
 formation in the disk of $7M_{\odot}pc^{-2}$, whereas for 
the halo phase we have several options with and without threshold. 
The threshold in the halo is $4M_{\odot}pc^{-2}$.
We also assume a constant surface mass density for the halo or variable with 
galactocentric distance. 
This is important for the gradients at very large galactocentric distances, 
where the enrichment from the halo predominates over the enrichment occurring 
in the thin disk thus influencing the abundances at such large distances.

We conclude that to reproduce the flat gradients suggested by the abundance measurements
at large galactocentric distances, we need to assume a constant density distribution
and a threshold in the star formation during the halo phase.
However, there are still many uncertainties in the data at very large
galactocentric distances and only more data will allow us to draw firm 
conclusions on this important point.

\item The chosen nucleosynthesis prescriptions (empirical yields
by Fran\c cois et al 2004) successfully reproduce
the abundances gradients  of each specific element, 
besides reproducing the [el/Fe] vs [Fe/H] relations
in the solar neighborhood, as shown already by Fran\c cois et al. (2004).

\item We also presented new results concerning the [La/Fe] vs. [Fe/H] relation
in the solar neighbourhood.
The data are the new ones by Fran\c cois et al. (2006). 
We conclude that La has the same origin as Ba: the bulk of La originates 
from low mass stars in the range 1-3$M_{\odot}$ as an s-process elements, 
but a fraction of  La originates, as an r-process element, from stars 
in the mass range 12-30$M_{\odot}$.

\end {itemize}

\section{Acknowledgments}

We thank Francesco Calura and Antonio Pipino
for several useful comments. G. C. and F. M.  acknowledge funds 
from MIUR, COFIN 2003, prot. N. 2003028039.
 C. C. acknowledges partial support by the INAF PRIN grant CRA 1.06.08.02.


\end{document}